\def\gs{\gamma_{\rm slow}}
\def\gf{\gamma_{\rm fast}}
\def\hks{h_{\rm KS}}
\def\ht{h_{\rm top}}
\def\lf{\lambda_F}

\documentstyle[preprint,aps]{revtex}

\begin{document}

\draft

\tighten

\preprint{\vbox{\hfill cond-mat/9605127 \\
          \vbox{\hfill UCSB--TH--96--08} \\ 
          \vbox{\hfill May 1996}
          \vbox{\vskip1.0in}
         }}

\title{The Onset of Chaos in the Quantum Hard-Sphere Gas}

\author{Mark Srednicki\footnote{E-mail: \tt mark@tpau.physics.ucsb.edu}}

\address{Department of Physics, University of California,
         Santa Barbara, CA 93106}

\maketitle

\begin{abstract}
\normalsize{
We show that the condition for the appearance of quantum chaos (Wigner-Dyson
distribution of energy eigenvalues, gaussian-random energy eigenfunctions)
in a dilute gas of many hard spheres is $\lambda \ll \ell$, where $\lambda$
is the wavelength of a typical particle and $\ell$ is the mean free path 
between collisions.  For fermions with Fermi wavelength $\lambda_F \ll \ell$,
this implies that all energy eigenstates, including the ground state, are 
chaotic.  Physical implications are discussed.
}
\end{abstract}


Quantum systems which are classically chaotic \cite{lh}
are known to exhibit several interesting and universal features,
such as Wigner-Dyson statistics for energy eigenvalues \cite{bgs}, 
gaussian-random energy eigenfunctions \cite{berry},
and quantum expectation values which approach classical microcanonical 
averages \cite{shnrl}.  
These features can then be used to understand certain physical
phenomena, such as the statistics of fluctuations in the conductance
of quantum dots \cite{ucf}.  
We will refer to these features collectively as ``quantum chaos''.
Quantum chaos is expected to appear only for sufficiently 
high energy eigenvalues; an important problem is the determination of
the energy scale which marks the (possibly gradual) onset of quantum chaos.
Here we consider this problem for a dilute gas of hard spheres,
a many-body system which is (almost certainly \cite{blp})
classically chaotic.

In general, the presence of quantum chaos at an energy $E$ requires 
two distinct conditions to be satisfied.  The first is that
\begin{equation}
\Delta \ll \hbar\gs \, ,
\label{1}
\end{equation}
where $\Delta$ is the mean spacing of energy eigenvalues near energy $E$,
and $\gs$ is the rate of the slowest classical process
which is physically relevant.  
For a system with $n \ge 2$ degrees of freedom,
condition (\ref{1}) is satisfied in the formal limit of $\hbar \to 0$
with $E$ fixed, since in that limit $\Delta \propto \hbar^n$.
If, on the other hand, we consider $\hbar$ to be fixed,
then (\ref{1}) implies a lower bound on the energy 
where we can expect the signatures of quantum chaos to appear.
Condition (1) can be understood heuristically by considering the
corresponding time scales (e.g., \cite{heller}).  Any relevant
classical process must occur before the Heisenberg time 
$\tau_H = 2\pi\hbar/\Delta$;
this is because $\tau_H$ is the time at which all observables
begin to oscillate quasi-periodically, an inherently quantum mechanical
phenomenon which shuts off any later classical effects.

For example, for a single particle in a random potential in $d$ dimensions 
(representing, e.g., an electron in a disordered metal),
$\gs \sim D/L^2$ is the Thouless rate \cite{local}; 
here $D \sim \ell v$ is the classical diffusion
constant, $\ell$ is the mean free path,
$v = (2E/m)^{1/2}$ is the speed of the particle,
and $L \gg \ell$ is the linear size of the system.
For the Sinai billiard in $d$ dimensions,
in which a particle moves ballistically in a cubic box
with a single spherical scatterer in the center,
$\gs \sim v/\ell \sim a^{d-1}v/L^d$,
where $v$ is the speed of the particle,
$L$ is the linear size of the box,
$a$ is the radius of the spherical scatterer, and
$\ell \sim L^d/a^{d-1}$ is the mean free path 
between collisions of the particle with it;
here we have assumed $a \ll L$, and hence $L \ll \ell$.
In both examples, $\Delta \sim (\lambda/L)^d E$, where 
$\lambda = 2\pi\hbar/mv$ is the wavelength of the particle.

For the example of a particle in a random potential, 
condition (1) leads to 
$\lambda^{d-1} \ll \ell L^{d-2}$.  
When $d=1$, this condition is independent of $\lambda$, and then it 
is not satisfied (since, by assumption, $\ell \ll L$ in this case).
For a particle in a random potential,
quantum chaos corresponds to the existence of extended 
(rather than localized) energy eigenstates \cite{local},
so (\ref{1}) correctly predicts that there are no extended states
when $d=1$.  It will turn out that the second condition, 
specified by Eq.\ (\ref{2}) below, is more stringent than (\ref{1}) 
when $d \ge 2$, and so we defer further discussion of this example.

For the example of the Sinai billiard, (\ref{1}) leads to
$\lambda \ll a$.  This condition is also necessary in order
to have large phase shifts when a particle with wavelength $\lambda$
is scattered from a sphere of radius $a$ in free space.
The appearance of quantum chaos only for $\lambda \ll a$
is also supported by some numerical computations \cite{mk}.

The second condition which must be satisfied 
for the appearance of quantum chaos is
\begin{equation}
\hbar\gf \ll E \, ,
\label{2}
\end{equation}
where $E$ is the total energy of the system and $\gf$ 
is the rate of the fastest classical process which is
physically relevant.  Like (\ref{1}), this condition is
satisfied in the formal limit of $\hbar \to 0$ with $E$ fixed,
and like (\ref{1}) it can be understood
heuristically by considering the corresponding time scales.
For a quantum state whose energy uncertainty is no larger than
the energy itself, $\hbar/E$ represents the minimal time uncertainty 
in any physical process.  A classically relevant process which takes
less time than $\hbar/E$ will not be reflected in the quantum theory.

In both of our examples, $\gf \sim v/\ell$ is the inverse of the
mean free time; condition (\ref{2}) then implies
$\lambda \ll \ell$.  In the example of the Sinai billiard, this
is weaker than the condition
$\lambda \ll a$ which follows from (\ref{1}).
In the example of a particle in a random potential, 
$\lambda \ll \ell$ is well-known from the theory of
weak localization as a necessary condition 
for the existence of extended states 
when $d \ge 3$ (e.g., \cite{weak}).
The critical case of $d=2$ is more complicated: 
all energy eigenstates are localized, but the
localization length is given by $\xi \sim \exp(c\,\ell/\lambda)\ell$,
where $c$ is a numerical constant.  If $\xi \gg L$, then
the states are effectively extended and the signatures
of quantum chaos will be present.

The most rigorous derivation of either (\ref{1}) or (\ref{2})
follows from the recent work of 
Andreev, Agam, Simons, and Altshuler (AASA) \cite{aasa},
who computed statistical properties of energy levels from
first principles, using only generic classical features
of chaotic systems, and making reasonable (but not strictly proven)
assumptions about the form of the classical limit of certain
functional integrals.  It is clear that the approach of AASA
extends in a straightforward way to properties of eigenfunctions
and matrix elements.  For a classical system with a Perron-Frobenius
spectrum (that is, the eigenvalues of the classical time evolution operator
when it is restricted to distributions on phase space that are suitably
well-behaved at all times) that has
a gap between the lowest eigenvalue of zero and the first nonzero
eigenvalue $\gamma_1$, AASA find that 
\begin{equation}
\gs = \mathop{\rm Re}\gamma_1 \, .
\label{gs}
\end{equation}
For a system with no gap, $\gamma_1$ can be identified with the
inverse of the time $\tau_1$ at which classical correlation functions
begin to decay as a power of the time (instead of exponentially).
The heuristic estimates of $\gs$ which we
made in our two examples are consistent with Eq.\ (\ref{gs}).

The value of $\gf$ in the AASA formalism follows
from the requirement that the Weyl symbol of a product of
two operators is equal (approximately) to the product of
their individual Weyl symbols.  In particular, this must
apply to products of the hamiltonian with itself.  Given
two operators $A$ and $B$ with Weyl symbols $A_W(x)$ and $B_W(x)$,
where $x=(q,p)$ are canonical coordinates on phase space and
\begin{equation}
A_W(x) = \int d^d q' \,
              e^{-ipq'/\hbar} \langle q+{\textstyle\frac12}q' |A| 
                                      q-{\textstyle\frac12}q' \rangle \, ,
\label{aw}
\end{equation}
the Weyl symbol of their product $AB$ can be expressed as \cite{om}
\begin{equation}
(AB)_W(x) = \exp\left[(i\hbar/2)
                      \nabla_1 \!\cdot\! J \!\cdot\! \nabla_2
                \right]
            A_W(x_1)B_W(x_2)|_{x_1=x_2=x} \, , 
\label{ab}
\end{equation}
where $J=((0,-I),(I,0))$ is the simplectic matrix.  To find $\gf$,
we must compute the discrepancy between $(H^2)_W(x)$ and $[H_W(x)]^2$,
and demand that it be small.  We begin by recalling that,
in a chaotic system, unstable periodic orbits are dense in phase space,
and that the number of them with period $T$ is proportional to 
$\exp(\ht T)$, where $\ht$
is the topological entropy.  Given an unstable periodic orbit, we can
choose phase space coordinates $x=(E,t,X)$ where
$E$ is the energy, $t$ measures length along the orbit,
and $X$ represents coordinates in the Poincar\'e section of the orbit.
For small $X$ we can write the Weyl symbol of the hamiltonian as
\begin{equation}
H_W(x) = E + \textstyle{\frac12} X \!\cdot\! \Lambda 
                                   \!\cdot\! X         \, ,
\label{hw}
\end{equation}
where $\Lambda$ is a diagonal matrix whose entries are the 
Lyapunov exponents $\lambda_i$, which are real and come in pairs
with equal magnitudes and opposite signs.  Applying (\ref{ab}), we find
\begin{equation}
(H^2)_W(x) = [H_W(x)]^2 - {\textstyle \frac14 \hbar^2 \sum_i \lambda_i^2}
             + O(\hbar^4)\, .
\label{h2w}
\end{equation}
Thus we must have $\hbar^2 \sum_i \lambda_i^2 \ll E^2$
in order for the symbolic manipulations of AASA to be valid.  

A similar but somewhat stronger condition follows from a closely related
point of the AASA analysis; given any function on phase space $f_W(x)$
which is smooth on the scale set by $\hbar$, AASA assume that
$(Hf)_W(x)$ is approximately equal to $H_W(x)f_W(x)$.
Let us consider the specific case of
\begin{equation}
f_W(x) = 1 + (\alpha/\hbar)X \!\cdot\! K \!\cdot\! X + \ldots \, ,
\label{fw}
\end{equation}
where $K = ((I,0),(0,-I))$, and $\alpha$ is a numerical constant
(which must be small if $f_W(x)$ is to be slowly varying).
We also write $\Lambda=((\Lambda^+,0),(0,-\Lambda^+))$, 
where $\Lambda^+$ is a diagonal matrix whose entries are 
the positive Lyapunov exponents $\lambda_i^+$.
Then we find
\begin{equation}
(Hf)_W(x) = H_W(x)f_W(x) 
            + {\textstyle \frac12 \alpha \hbar \sum_i \lambda_i^+}
            + O(\hbar^2)\, .
\label{hf}
\end{equation}
Thus we must have $\hbar \sum_i \lambda_i^+ \ll E$.
The sum of the positive Lyapunov exponents is
the Kolmogorov-Sinai entropy $\hks$,  
and so we can make the identification
\begin{equation}
\gf = \hks \, .
\label{gf}
\end{equation}
Of course, by choosing other forms for the matrix $K$, we can equally
well conclude that we must have $\hbar\sum_i \alpha_i\lambda_i \ll E$
for any coefficients $\alpha_i$ which are each of order unity or smaller.
The Kolmogorov-Sinai entropy $\hks$ is simply the largest linear
combination of this type.

Eq.\ (\ref{gf}) also follows from an analysis of
the validity of the ``diagonal approximation'' \cite{diag}
to the semiclassical periodic orbit expansion of, e.g., the spectral density
correlator.  When valid, the diagonal approximation can be used to derive
some of the signatures of quantum chaos \cite{perorb}.
Without employing any resummation techniques, 
the periodic orbit expansion
converges only if the energy is given an imaginary part
whose magnitude exceeds $\hbar(\ht - \frac12 \hks)$ \cite{imag}.
When $E$ is kept strictly real, the diagonal approximation
relies on large $E$ to produce rapidly varying phases which
then yield large cancellations.  These can be expected to occur only if 
$E \gg \hbar(\ht - \frac12 \hks)$.  Since in general 
$\ht \ge \hks$, we once again find the condition 
$\hbar\hks \ll E$.

We now turn to our main subject, the application of 
(\ref{1}) and (\ref{2}) to a gas of $N \gg 1$ hard spheres,
each with radius $a$, in a box with edge length $L$ in $d$ dimensions.
We will take the gas to be dilute, so that $N a^d \ll L^d$.
The length of the mean free path between collisions of one
particular sphere with any other is then
$\ell \sim L^d/Na^{d-1}$, which is much larger than $a$;
we will, however, assume that $\ell \ll L$.

We must now determine, as functions of the total energy $E$,
the key quantities $\gs$, $\gf$, and $\Delta$.
The low lying modes of the Perron-Frobenius operator
are associated with diffusive processes, and so we expect
$\gs \sim D/L^2 \sim \ell v/L^2$,
where $v = (2mE/N)^{1/2}$ is the typical speed of a single particle.
Thus, in the thermodynamic limit of $N\to\infty$ with $E/N$ and $L^d/N$
fixed, $\gs$ goes to zero like $N^{-2/d}$.
However, in a many-body system, the level spacing $\Delta$ 
is related to the thermodynamic entropy $S$ via $\Delta \sim e^{-S}E$.
Since entropy is an extensive quantity,
$\Delta$ is exponentially small in $N$ as $N\to\infty$.
Therefore condition (\ref{1}) is always satisfied for large $N$,
independent of $E$.

To evaluate condition (\ref{2}), we need
the Kolmogorov-Sinai entropy of the hard-sphere gas;
it is given by \cite{gasp}
\begin{equation}
\hks \sim (Nv/\ell)\ln(\ell/a) \, ,
\label{hks}
\end{equation}
where again $v = (2mE/N)^{1/2}$ is the typical speed of a single particle.
If define the single-particle wavelength $\lambda = 2\pi\hbar/mv$,
we then find that condition (\ref{2}) yields
\begin{equation}
\lambda \ll {\ell \over \ln(\ell/a)} \, .  
\label{cond}
\end{equation}
This, then, is the requirement for the
appearance of quantum chaos in a dilute gas of many hard spheres.

Let us eliminate a possible loophole in the preceding argument.
We have assumed that the spectrum of single-particle wavelengths
contained in an energy eigenstate is characterized by a single
length scale $\lambda \sim (\hbar^2N/mE)^{1/2}$.  
Fortunately, in the energy range
where the signatures of quantum chaos appear, this is correct:
the distribution of single-particle wavelengths in a chaotic energy 
eigenstate is thermal, with the characterizing temperature
related to the energy eigenvalue by the usual thermodynamic
relation between temperature and energy \cite{sred}.  
Of course, which thermal distribution we get 
(Maxwell-Boltzmann, Bose-Einstein, Fermi-Dirac)
depends on the assumed statistics of the particles, an issue we have
ignored up to now.

Here we will consider the case of fermions, which, for sake of
definiteness, we will assume to have spin one-half.  (Spin degeneracy
will, however, affect only various numerical factors which we will
not evaluate explicitly.)  The Weyl symbol calculus can be modified
to reflect Fermi statistics \cite{gn}, 
as can the periodic orbit expansion \cite{gasp,ce}.
Therefore all of our previous analysis is still applicable.
The ground state of the system is characterized by a Fermi wavelength
$\lf \sim L/N^{1/d}$; note that, for a dilute gas, $a \ll \lf \ll \ell$.
Since $\lf \ll \ell$, the ground state (and, indeed,
every energy eigenstate) apparently meets our test
for the appearance of quantum chaos.

Our claim of quantum chaos in the ground state 
must be reconciled with the standard theory of the
normal Fermi liquid \cite{lp}, in which quasiparticle excitations
have weak couplings characterized by the small parameter
$a/\lf$.  The key point is that quasiparticles do not
represent energy eigenstates; this is clear from the facts
that quasiparticle energies have an imaginary part,
and that their number is not conserved.  Instead,
quasiparticles represent superpositions of energy
eigenstates with particularly simple behavior under
time evolution.  A quasiparticle is analogous to a wave packet
in the Sinai billiard, narrow in both position and momentum,
constructed by appropriate superposition of high energy
eigenfunctions.  Such a packet will move simply (that is,
ballistically) for a long time.  This is true even though
the individual energy eigenfunctions of which it is composed
are themselves chaotic.
The difference from the case of the normal Fermi liquid
is that there we do not require high energies
for the occurrence of quantum chaos.

Another point is that the results of diagrammatic perturbation theory
(for the ground state expectation values of simple operators) imply that
the exact ground-state wave function can be written
as an asymptotic series of the form
\begin{equation}
\psi_0(q) = \psi_0^{(0)}(q) + (a/\lf) \psi_0^{(1)}(q) + \ldots \, ,
\label{psi}
\end{equation}
where $\psi_0^{(0)}(q)$ is the usual Slater determinant of eigenfunctions
of a single particle in a box.  Quantum chaos implies (among other things)
that the distribution of the values of $\psi_0(q)$ over the 
configuration space is gaussian \cite{berry,mk,gauss}.  
This can be reconciled with
(\ref{psi}) only if the same is true of the zeroth-order approximation
$\psi_0^{(0)}(q)$.  A gaussian distribution of the values of 
$\psi_0^{(0)}(q)$ implies, and is implied by,
\begin{equation}
\int d^{Nd}q\,\left[\psi_0^{(0)}(q)\right]^k = 
       \cases{(k-1)!! & for $k$ even \cr
                    0 & for $k$ odd \cr} \, ,
\label{int}
\end{equation}
where we have set $L=1$ for notational simplicity.
Fortunately, Eq.\ (\ref{int}) can be shown to be valid for large $N$.
This is more easily demonstrated if we put periodic boundary conditions
on the box (instead of hard walls) so that the single-particle 
eigenfunctions are plane waves. Then, if we keep only those terms
on the left-hand side of (\ref{int}) in which each single-particle
eigenfunction $\chi_{\bf p}({\bf x}) = e^{i{\bf p\cdot x}/\hbar}$
appears together with $\chi_{\bf -p}({\bf x})$, 
the result is the right-hand side of (\ref{int}).
The remaining terms on the left-hand side are
much more numerous, but most of them integrate to zero, and
those that do not can have either sign.  Assuming that these
signs can be treated as statistically random, one can show
that the net contribution of the remaining terms to the 
right-hand side of (\ref{int}) is exponentially small in $N$. 

The appearance of quantum chaos at all energies in the
normal Fermi liquid does have physical consequences
for quasiparticle behavior on long time scales.
It predicts that thermal equilibrium will eventually
be attained, even in the complete absence of any
coupling to an external environment, and beginning
from an arbitrary initial state \cite{sred}.
Thus quantum chaos can serve as the fundamental dynamical
underpinning of the usual phenomenological transport
equations (which also predict, in more detail, an approach to thermal
equilibrium).  In this regard, it is interesting to note that
the condition $\lambda \ll \ell$ also arises during a careful 
derivation of a Markovian transport equation for a dilute gas with 
a smooth, short-range interaction potential between particles \cite{rm}.
(This formalism, however, is applicable only to a highly restricted
class of initial states.)

As another possible physical consequence, let us note that
a dilute gas of hard-sphere fermions in a box is a poor model
of a nucleus.  Nevertheless, if we extrapolate the present result
to a dense gas with a weak long range attraction
(replacing the box) but a short range repulsion, 
it may be that quantum chaos also extends
down to the ground state in large nuclei \cite{boh}.

To conclude, we have explored the conditions for the appearance
of quantum chaos in a dilute gas of hard spheres, and found that
they imply that a typical particle's wavelength $\lambda$
must be much less that the classical mean free path $\ell$.
This condition is satisfied by all energy eigenstates,
including the ground state, if the particles are fermions.
This result is consistent with the standard theory of the
normal Fermi liquid.  It can help to put phenomenological
transport equations on a firmer theoretical foundation,
and may have implications for the properties of low-lying
energy eigenstates in large nuclei.

\begin{acknowledgments}
I am happy for the opportunity to thank
O.~Agam, N.~Argaman, O.~Bohigas, M.P.A.~Fisher, S.~Fishman, I.~Guarneri,
and J.S.~Langer for helpful discussions.
\end{acknowledgments}

\end{document}